\magnification=\magstep1
\baselineskip=24 true pt
\hoffset=0.0 true cm
\voffset=0.5 true cm
\hsize=17.0 true cm
\vsize=21.0 true cm
\parindent=30 true pt
\parskip=9 true pt
\raggedbottom


\centerline{\bf ON THE PARAMETERS OF LEWIS METRIC}
\centerline{\bf FOR THE WEYL CLASS}
\bigskip
\bigskip
\bigskip
{\baselineskip =14 true pt \centerline{\bf M. F. A. da Silva}
\centerline{\tenit Departamento de F\'{\i}sica e Qu\'{\i}mica, Universidade 
Estadual Paulista,}
\centerline{\tenit Av. Ariberto Pereira da Cunha 333, 12500, 
Guaratinguet\' a - SP, Brazil and}
\centerline{\tenit Departamento de Astrof\'{\i}sica, CNPq-Observat\'orio 
Nacional,}
\centerline{\tenit Rua General Jos\' e Cristino 77, 20921-400, Rio de Janeiro 
- RJ, Brazil.}}
\bigskip
{\baselineskip =14 true pt \centerline{\bf L. Herrera}
\centerline{\tenit Departamento de F\'{\i}sica, Facultad de Ciencias,} 
\centerline{\tenit Universidade Central de Venezuela and} 
\centerline{\tenit Centro de F\'{\i}sica, Instituto Venezolano de 
Investigaciones Cient\'{\i}ficas,}
\centerline{\tenit Caracas, Venezuela. Postal address: Apartado 80793, 
Caracas 1080A, Venezuela}}
\bigskip
{\baselineskip =14 true pt \centerline{\bf F. M. Paiva and N. O. Santos}
\centerline{\tenit Departamento de Astrof\'{\i}sica, CNPq-Observat\'orio 
Nacional,}
\centerline{\tenit Rua General Jos\'e Cristino 77, 20921-400, Rio de Janeiro
 - RJ, Brazil}} 

\bigskip
\bigskip
\noindent Author's internet addresses respectively: mfas@on.br,
lherrera@conicit.ve, fmpaiva@on.br and nos@on.br. 

\vfill\eject

\centerline {\bf Abstract}
We provide physical interpretation for the four parameters of the
stationary Lewis metric restricted to the Weyl class. Matching this
spacetime to a completely anisotropic, rigidly rotating, fluid
cilinder, we obtain from the junction conditions that one of these
parameters is proportional to the vorticity of the source. From the
Newtonian approximation a second parameter is found to be proportional
to the energy per unit of length. The remaining two parameters may be
associated to a gravitational analog of the Aharanov-Bohm effect. We
prove, using the Cartan scalars, that the Weyl class metric and static
Levi-Civita metric are locally equivalent, i.e., indistinguishable in
terms of its curvature.
\vfill\eject

{\bf1 Introduction}

\noindent Lewis stationary vacuum metric [1] is usually presented with
four parameters [2] without being given a specific physical
interpretation unless matched to a particular source. Here we shall
provide physical interpretation for the four parameters characterizing
the Lewis metric, when the source consists of a general rigidly
rotating anisotropic fluid cylinder.

The Lewis metric solution of Einstein's equations can be split into two 
families called Weyl class and Lewis class. Here we shall restrict our 
study to the Weyl class where all four parameters appearing in the metric 
are real. For the Lewis class these parameters can be complex. 

Matching the Weyl class metric to a cylindrical fluid, from the
junction conditions one of the parameters is found to be proportional
to the vorticity of the source.  Examination of the Newtonian limit
shows that a second parameter is proportional to the energy per unit
length, also in the same limit a third parameter appears as the
arbitrary constant potential which is always present in the Newtonian
solution. Finally the fourth parameter, together with the first, is
found to be responsible for the non-staticity of Lewis metric and has
to vanish when the vorticity of the source tends to zero if one wishes
to recover the Levi-Civita static vacuum metric [3], in that limit.
Additionally these two last parameters may be related to a
gravitational analog of the Aharanov-Bohm effect.

The physical properties of solutions of the Einstein's equations are
not obvious from a given metric tensor, in part due to the coordinate
arbitrariness in general relativity. Using an approach borrowed from
the equivalence problem [4,5], and based on the Cartan
scalars, we shall study the local properties of the Lewis spacetime.
It is demonstrated that the Weyl class metric is indistinguishable of the 
Levi-Civita metric from the curvature point of view. However we show that 
they are topologically different.

The paper is organized as follows: in the next three sections we give
the notation and the spacetime at both sides of the boundary of the
source. In section 5 we introduce the Cartan scalars technic and apply
it to investigate the local properties of the Lewis metric. In section
6 the junctions conditions are discussed. The Newtonian approximation
is presented in section 7, together with  a discussion about the
gravitational Aharanov-Bohm effect. In section 8 we present a conclusion. 
Three appendixes are added, at the end, for the Ricci and Riemann tensor 
components and the algebraic invariants of the Riemann tensor. These 
quantities are calculated for the metric that we shall study.

\bigskip
\bigskip
\bigskip
{\bf 2 Spacetime}

\noindent The spacetime we describe by the general cylindrically 
symmetric stationary metric
$$ds^2=-f dt^2 +2kdtd\varphi +e^{\mu} \left( dr^2 +dz^2 \right)  
+ld\varphi^2,\eqno(2.1)$$ 
where $f$, $k$, $\mu$ and $l$ are functions only of 
$r$, and the ranges of the coordinates $t$,$z$ and $\varphi$ are $$-\infty 
<t<\infty,\qquad -\infty<z<\infty,\qquad 0\leq \varphi \leq
2\pi,\eqno(2.2)$$ the hypersurfaces $\varphi=0$ and $\varphi=2\pi$ being 
identified. The coordinates are numbered $$x^0=t,\qquad x^1=r,\qquad x^2=z,
\qquad x^3=\varphi.\eqno(2.3)$$
The metric $(2.1)$ has to satisfy Einstein's field equations, 
$$R_{\mu\nu}=\kappa \left( T_{\mu\nu}-{1\over 2}
g_{\mu\nu}T \right),\eqno(2.4)$$ given in its usual
 notation. The non zero components of the Ricci tensor
$R_{\mu\nu}$ for $(2.1)$ are given
in the appendix A.

The spacetime is devided into two regions: the interior, with $0\leq r 
\leq R$, to a cylindrical $\Sigma$ surface of radius $R$ centered along $z$; 
and the 
exterior, with $R \leq r < \infty$. 
\bigskip
\bigskip
\bigskip
{\bf 3 Exterior spacetime}

\noindent The exterior spacetime is filled with vacuum, hence Einstein's 
equations 
$(2.4)$ reduce to $$R_{\mu\nu}=0. \eqno (3.1)$$ The general solution of $(3.1)$
for $(2.1)$ is the stationary Lewis metric [1] given in reference [2] notation, 
$$f=ar^{-n+1} - 
{c^2\over{n^2 a}}r^{n+1},\eqno(3.2)$$ $$k=-Af,\eqno(3.3)$$ $$e^\mu=r^{{1\over 
2}\left(n^2-1\right)},\eqno(3.4)$$ $$l={r^2 \over f} -A^2 f,\eqno (3.5)$$ with 
$$A={cr^{n+1}\over {naf}} + b. \eqno(3.6)$$
The constants $a$, $b$, $c$ and $n$ can be either real or complex, the 
corresponding solutions belong to the Weyl class or Lewis class, 
respectively. Here we restrict our study to the Weyl class. 

The transformation [7]
$$d\tau=\sqrt{a}(dt+b d\varphi),\eqno (3.7)$$ 
$$d\bar \varphi={1\over n}[-c dt+(n-bc) d\varphi],\eqno (3.8)$$ 
casts the metric (2.1) into
$$ds^2 = -r^{-n+1}d\tau^2 + 
r^{{1\over2}\left(n^2-1\right)}\left(dr^2 + dz^2\right)+{1\over a} r^{n+1} 
d\bar \varphi ^2.\eqno(3.9)$$
This is locally the Levi-Civita metric. Nevertheless, since $\varphi=0$
and $\varphi=2\pi$ are identified, $\tau$ defined in (3.7) attains a
periodic nature unless $b=0$ [7]. On the other hand, the new coordinate
$\bar \varphi$ ranges now from $-\infty$ to $\infty$. A more detailed
accounting on this subject can be found in Stachel [8,9]. In order to
globally transform the Weyl class of the Lewis metric into the static
Levi-Civita metric we have to make $b=0$ and $c=0$. Note that in this
case, from the transformations (3.7) and (3.8), $\tau$ and $\bar
\varphi$ become respectively true time and angular coordinates. 

Hence we can say that $b$ and $c$ are responsible for the 
non-staticity of 
this family of solutions in the Lewis metric.

\bigskip
\bigskip
\bigskip
{\bf4 Interior spacetime}

\noindent The interior spacetime is described by a cylinder filled with 
stationary aniso-\break tropic fluid, hence the energy momentum tensor 
$T_{\mu\nu}$ is given 
by $$T_{\mu\nu}=(\varrho +P_r)u_{\mu} u_{\nu} + P_r g_{\mu\nu} +
(P_{\varphi} -P_r)K_{\mu} K_{\nu} + (P_z - P_r)S_{\mu} 
S_{\nu},\eqno(4.1)$$ where $\varrho$ is the energy density, $P_r$, $P_z$ and 
$P_\varphi$ are the principal stresses, and $u_\mu$, $K_\mu$ and $S_\mu$ 
are four vectors satisfying $$u_{\mu} u^{\mu} =-1,\qquad K_{\mu} 
K^{\mu}= S_{\mu} S^{\mu} =1,\eqno(4.2)$$ $$K_{\mu} u^{\mu} = S_{\mu} 
u^{\mu} = K_{\mu} S^{\mu} =0.\eqno (4.3)$$ Considering the source under 
rigid rotation and choosing the coordinates in a comoving frame, we obtain 
for $u^\mu$, $K^\mu$ and $S^\mu$ using $(2.1)$ and $(4.2-3)$ $$u^\mu = 
\left({1\over\sqrt{f}},0,0,0 \right),\eqno(4.4)$$ 
$$K^\mu = \left( {k\over{D\sqrt{f}}}, 0, 0,
{\sqrt{f}\over D} \right),\eqno(4.5)$$ $$S^\mu = \left(0,0,e^{-{\mu\over 2}}, 
0 \right).\eqno(4.6)$$ Calculating the energy 
momentum 
tensor $(4.1)$ using $(2.1)$ and $(4.4-6)$ we obtain that the non null 
components are 
$$T^0_0 = -\varrho,\qquad 
T^1_1 = P_r,\qquad T^2_2 = P_z,\qquad T^3_3 = P_\varphi,\qquad T^0_3 = 
{k\over f}(\varrho +P_\varphi).\eqno(4.7)$$ We observe that $T^3_0 = 0$, 
hence the corresponding component of Einstein's field equations $(2.4)$ 
becomes $$\left( {fk^\prime - kf^\prime \over D} \right)^\prime = 0,
\eqno(4.8)$$ where we 
used $(A.2)$. Integrating $(4.8)$ we obtain $$fk^\prime - kf^\prime = \alpha 
D, \eqno(4.9)$$ 
where $\alpha$ is an integration constant. 

The vorticity tensor $\omega_{\alpha\beta}$ of the matter inside the
cylinder is given by $$\omega_{\alpha\beta} = {1\over 2}
(u_{\alpha;\beta} - u_{\beta;\alpha})+{1\over 2}(u_{\alpha;\nu}u^\nu
u_\beta-u_{\beta;\nu}u^\nu u_\alpha).\eqno(4.10)$$ With $(2.1)$ and
$(4.4)$ we obtain the non null components of $(4.10)$
$$\omega_{31}={1\over 2}\left({k^\prime\over{\sqrt{f}}}-{kf^\prime\over
f\sqrt{f}} \right).\eqno(4.11)$$ The vorticity vector $W^\alpha$ is
$$W^\alpha={1\over2}{\varepsilon^{\alpha \beta \gamma \delta}\over
\sqrt{-g}}u_\beta \omega_{\gamma \delta},\eqno(4.12)$$ and with
$(2.1)$, $(4.4)$ and $(4.11)$ we obtain $$W^\alpha={1\over2}
\varepsilon^{3\alpha 10}{1\over D e^\mu} \left(k^\prime-k{f^\prime\over
f} \right).\eqno(4.13)$$ Now calculating the magnitude of $W^\alpha$
from $(4.13)$ and considering $(4.8)$ we have $$\left(W_\alpha W^\alpha
\right)^{1\over2}={\xi \over2fe^{{\mu \over2}}}.\eqno(4.14)$$ Hence we
see from $(4.14)$ that $\xi$ measures the vorticity of the cylinder.

Althought we did not solve all the interior field equations for the
general source (4.1) it can be easily shown that solutions exist. In
fact Krasi\' nski [10, 11, 12] found a complete set of solutions of the
field equations for a perfect fluid source and matched his solutions to
the corresponding exterior solutions [13]. Another and well known
interior solution is that due van Stockum [14], with vanishing
pressure. In this particular case the expression for the vorticity
(4.14) reduces to the correspondent Bonnor result [7] where his $a$ is
equal $-{\xi\over 2}$. 
\bigskip
\bigskip
\bigskip
{\bf 5 Cartan scalars}

\noindent In this section, the equivalence problem and its solution due to 
Cartan [4] and Karlhede [5] are presented. We show how
it leads to the Cartan scalars and to a local characterization of
spacetimes. At the end of this section we apply this approach to the Weyl 
class metric (3.2-5).

The equivalence problem consists of deciding whether a coordinate
transformation $\tilde{x}^a = \tilde{x}^a(x)$ exists which takes a
metric $g_{ab}(x)$ into another metric $\tilde{g}_{cd}(\tilde{x})$. If
it exists, these two metrics are called equivalent. A first attempt to
solve this problem would probably be to use the scalars made by
contraction of the Riemann tensor (the so called 14 algebraic invariants) 
and its
covariant derivatives.  Unfortunately, this fails, i.e., two metrics
may have the same scalars (built in the above way) and be not
equivalent. For example, these scalars are all zero for plane-waves
and Minkowski spacetimes and yet they are not equivalent metrics
[15]. We present in appendix C the non null 14 algebraic invariants of 
the Riemann 
tensor calculated for the metric (3.2-5) if the reader wants to compare 
them with the Cartan scalars obtained below. 
The best known solution to the equivalence problem was
presented by Cartan [4] and can be summarized as follows.

Let $\omega^A~(A = 0,\ldots,3)$ be a frame such that the line element
can be written as $ds^2 = \eta_{AB} ~\omega^A ~\omega^B$. The
components of the Riemann tensor and its covariant derivatives (up to
possibly the $10^{\rm th}$ order) calculated in a constant frame
($\eta_{AB}$ constant) shall be called {\it Cartan scalars}. This name
comes from the fact that the Cartan scalars transform like scalars under
coordinate transformations. Notice, however, that they transform like
tensor components under frame transformations. With this in mind, we
can state the equivalence theorem, due to Cartan [4] as:

\vskip .5cm

Theorem: {\it Two metrics are equivalent if and only if there exist
coordinate and Lorentz transformations which transform the Cartan
scalars of one of the metrics into the Cartan scalars of the other.}

\vskip .5cm

As introduced above, the Cartan scalars are defined up to Lorentz and
coordinate transformations. Lorentz transformations present no great
problem since the frame may be uniquely fixed (up to isotropies of the
metric) by choosing a standard form to the set of Cartan scalars. Once
a standard frame is chosen, the Cartan scalars become defined up to
coordinate transformations. Although there is no algorithm for deciding
whether there is a coordinate transformation which takes one set of
Cartan scalars into the other, it is, in most practical cases, a decidable
task.

The practical algorithm for calculating the Cartan scalars was
developed by  Karlhede [5]. Here we present a summary (for
recent reviews see, for instance
[5,16,17]). Such an
algorithm could have, in principle, 11 steps, from the $0^{\rm th}$
order derivative up to the $10^{\rm th}$. This number can be thought of
as related to the 6 Lorentz transformations, the 4 coordinates of the
spacetime and one integrability condition. At each order of derivation,
the frame is fixed up to the isotropies of the Cartan scalars
calculated until this order. At each order, the number of the
functionally independent functions of the coordinates among this set of
Cartan scalars is also found. How to fix the frame and how to find the
independent functions are explained in, for instance,
[16,18]. There will be a moment when no new
information for fixing the frame and no new functionally independent
function of the coordinates are found. At this point, no new derivatives
are necessary [4] and the set of Cartan scalars is complete, in
the sense that they are necessary and sufficient for deciding the
equivalence problem.

Since the Cartan scalars are sufficient for deciding about the
equivalence of metrics, they contain all local information about a
metric and provide a local characterization of a spacetime.
Nevertheless, global properties, such as topological deffects, would
probably not appear in the Cartan scalars.

The Cartan scalars can be calculated by using the computer algebra
system SHEEP [19], the program CLASSI [20] and their interface with REDUCE 
[15]. The practical implementation 
of the algorithm works with
spinors rather than tensors, since symmetries are better handled in this
way.  MacCallum and {\AA}man [21] found a minimal set of
algebraically independent Cartan scalars in the spinorial formalism
which, for vacuum solutions, as is the case of the Weyl class metric, are,
until the first order:
$$0^{\rm th}~{\rm derivative}: \Psi_A,~A = (0,\ldots,4); $$
$$1^{\rm st}~{\rm derivative}: 
\nabla\Psi_{A B^{'}},~A = (0,\ldots,5),~B = (0,~1); $$
where $\Psi_A$ is the Weyl spinor and $\nabla\Psi_{A B^{'}}$ is the first 
symmetrized covariant derivatives of $\Psi_A$ [15].

In order to calculate the Cartan scalars for the Weyl class metric (3.2-5), 
we chose the following Lorentz tetrad:

$$\eqalign{ \theta^0 & = \sqrt{f}dt -{k\over{\sqrt{f}}}d\varphi, 
\cr
\theta^1 & = {1\over{\sqrt{f}}}rd\varphi ,\cr
\theta^2 & = e^{\mu\over 2}dr, \cr
\theta^3 & = e^{\mu\over 2}dz. \cr}\eqno(5.1)$$
Since the Cartan scalars are calculated in the spinorial formalism, a
null frame must be used, which we define from the above by
$$\eqalign{ \omega^0 & = {1\over \sqrt{2}}(\theta^0 +   \theta^1), \cr
\omega^1 & = {1\over \sqrt{2}}(\theta^0 -   \theta^1), \cr
\omega^2 & = {1\over \sqrt{2}}(\theta^2 + i \theta^3),  \cr
\omega^3 & = {1\over \sqrt{2}}(\theta^2 - i \theta^3).  \cr} \eqno(5.2)$$

Following the usual technics [5], we find that the standard null frame
is given by the transformation
$$\eqalign{ \tilde{\omega}^0 & =\Bigl( {{na-cr^n}\over {na+cr^n}}\Bigr) 
^{1\over 2}
\omega^0, \cr
\tilde{\omega}^1 & = \Bigl( {{na+cr^n}\over {na-cr^n}}\Bigr) ^{1\over 2} 
\omega^1 , \cr
\tilde{\omega}^2 & = \omega^2 , \cr
\tilde{\omega}^3 & = \omega^3 . \cr} \eqno(5.3)$$
The correspondent standard Lorentz frame can be found from this one using 
(5.2), which gives:
$$\eqalign{ \tilde{\theta}^0 & =
na\Bigl( {f\over {n^2 a^2 -c^2 r^{2n}}} \Bigr) ^{1\over 2} (dt+bd\varphi ),
\cr
\tilde{\theta}^1 & = 
\Bigl( {f\over {n^2 a^2 -c^2 r^{2n}}} \Bigr) ^{1\over 2}  r^n
\left\lbrack -cdt+(n-bc)d\varphi \right\rbrack , \cr
\tilde{\theta}^2 & = e^{\mu \over 2}dr, \cr
\tilde{\theta}^3 & = e^{\mu \over 2}dz. \cr} \eqno(5.4)$$

In the standard null frame, the Cartan scalars are found to be
$$\Psi_2=-{1\over 8}(n^2-1)r^{-{1\over 2}(n^2+3)},
\eqno(5.5) $$
$$\Psi_0 =\Psi_4=-n\Psi_2 ,\eqno(5.6) $$
$$\nabla\Psi_{01'} =\nabla\Psi_{50'}=-{\sqrt {2}\over {16}}n(n^4-1)
r^{-{3\over 4}(n^2+3)},\eqno(5.7)$$
$$\nabla\Psi_{10'}=\nabla\Psi_{41'}=-{\sqrt {2}\over 8}n(n^2-1)
r^{-{3\over 4}(n^2+3)},\eqno(5.8)$$
$$\nabla\Psi_{21'}=\nabla\Psi_{30'}={\sqrt {2}\over {32}}(n^2+3)(n^2-1)
r^{-{3\over 4}(n^2+3)}.\eqno(5.9)$$

Since no new information appears in the first derivative, it is not
necessary to find the second derivative. From the Cartan scalars, we
can find that the metric is Petrov type I and therefore has no isotropy
group and that it has a 3 dimensional isometry group, since the Cartan
scalars present a single functionally independent function of the
coordinates.

From the Cartan scalar we see that only the parameter $n$ helps to
curve the
spacetime for the Weyl class metric. Hence, it can be showed that the
Cartan scalars for the Levi-Civita metric (3.9) are identical to those
of the Weyl class of the Lewis metric. Therefore, from the curvature
point of view, Levi-Civita static metric (3.9) and the Weyl class
stationary metric are indistinguishable, which confirms the coordinate
analysis of the end of the section 3. However, as it will be shown
below, these two metrics posses very different topological behaviour.

\bigskip
\bigskip
\bigskip
{\bf 6 Junction conditions}

\noindent On the surface of discontinuity $\Sigma$, $r=R$, the junction 
conditions are those of Darmois [22], namely, that the first and second 
fundamental forms have to be continuous. Choosing the same coordinates for 
the exterior and interior spacetimes these conditions become
$$\left \lbrack g^-_{\mu \nu} -g^+_{\mu \nu} \right \rbrack _{\Sigma} =0,
\eqno(6.1)$$
$$\left \lbrack g^{-\prime} _{\mu \nu} -g^{+\prime} _{\mu \nu} \right 
\rbrack _{\Sigma} =0,\eqno(6.2)$$
where the indexes $-$ and + stand for the interior and exterior spacetimes 
respectively.

At $r=R$ considering (3.2-6), (4.9) and (6.1-2) we obtain
$$c=-{{\xi} \over 2}.\eqno(6.3)$$
Hence, from (6.3) and (4.14) we have that the constant $c$ in the Weyl 
class 
metric measures the
vorticity of the source described by a rigidly rotating anisotropic fluid 
cylinder. For the particular case of the null pressure Bonnor's result [7] 
is recovered.

\bigskip
\bigskip
\bigskip
{\bf 7 Newton limit and Ahanarov-Bohm effect}

\noindent In the Newtonian limit, the vorticity term is 
negligible, then from (6.3) $c\approx 0$, and 
calling $(3.2)$ 
$$f=e^{2U},\eqno(7.1)$$ 
then $U$ is 
$$U=2\sigma \ln{r}+{1\over 2} \ln{a}, \eqno(7.2)$$ 
where $\sigma$ is given by 
$$\sigma={1\over 4} \left( 1-n \right).\eqno(7.3)$$  
In Newton's theory, (7.2) is the gravitational potential of an infinite 
uniform line-mass with mass per unit length $\sigma$.
The constant ${1\over 2}\ln{a}$ in $(7.2)$ represents the constant 
arbitrary potential 
that exists in the Newtonian solution. Timelike circular geodesics exists 
for $1>n>0$, or $0<\sigma <{1\over 4}$, as expected in the Newtonian 
analog. When $\sigma ={1\over 4}$ the circular geodesics become null 
[6,23].

The metric (3.2-6) has infinite curvature, according to the Cartan scalars 
(5.5-8),
only at $r=0$ for all $n$ except $n=\pm 1$, i.e., $\sigma =0$ and $1\over 2$ 
according to (7.3). Thus the Weyl class metric has a singularity along the 
axis $r=0$, then we can say that this spacetime is generated by an 
infinite uniform line source for densities $0<\sigma <{1\over 4}$. 

Considering the static limit for the Weyl class metric when $n=1$ 
$\left(\sigma =0\right)$ and $b=c=0$ we have from (2.1) 
$$ ds^2=-dt^2 + dr^2 + dz^2 + {r^2\over a}d\varphi^2,\eqno(7.4)$$ 
giving rise to 
a locally flat spacetime. In this case (7.4) represents the spacetime
generated by a string along the axis of symmetry with linear energy density
$\lambda$ given by [25] 
$$\lambda={1\over 4} \left(1-{1\over \sqrt{a}} \right),\eqno(7.5)$$ 
being $a>1$. Hence the constant $a$ is directly 
linked to the gravitational analog of Aharanov-Bohm effect [26]. This effect
shows that gravitation depends on the topological structure of spacetime 
giving rise to an angular defficit $\delta$ equal to 
$$\delta=2\pi\left(1-{1\over \sqrt{a}}\right),\eqno(7.6)$$ 
as in the eletromagnetic Aharanov-Bohm effect, where a (classical) non-observable quantity 
(the vector potential) becomes observable (part of it) through a quantum non-local effect. Its 
gravitational analog allows a (Newtonian) non-observable quantity (the 
additional constant to the Newtonian potential) to become observable in the
 relativistic theory through the angular deficit in strings.

Considering $c=0$ and $n=1$ $\left(\sigma=0\right)$ in (2.1) we have from (3.2-6) and 
(3.8) 
$$ds^2 = -d\tau ^2 - 2b\sqrt{a} d\tau d\varphi +dr^2 +dz^2 +
\left({r^2\over a} - b^2 a \right)d\varphi^2,\eqno(7.7)$$ 
producing a locally flat 
spacetime. In this case (7.7) represents the exterior spacetime of a 
spinning string along the axis of symmetry [27] with  the same linear energy 
density $\lambda$ given by (7.5) and angular momentum $J$ given by 
$$ J=-{1\over 4}b\sqrt{a}, \eqno(7.8)$$ 
being $a>1$.

As it has been recently shown [27], a quantum scalar particle moving around
a spinning cosmic string as given by (7.7), exhibits a phase factor 
proportional to $J$, in its angular momentum. An evident reminiscence
 of the Aharanov-Bohm effect. It is also worth mentioning that even if $b=0$,
  an Aharanov-Bohm like effect (of a different kind) appears (as commented in 
  the static case), since the angular momentum spectrum differs from the
 usual one, if only $a>1$.

\bigskip
\bigskip
\bigskip
{\bf Conclusion}

\noindent We obtained physical interpretations for the four real parameters
$n$, $a$, $b$, $c$ appearing in the Lewis metric (3.2-5) for the Weyl 
class.

The parameter $n$ is associated to the Newtonian mass per unit length of an
uniform line mass when it produces low densities.

The parameter $a$ is connected to the constant arbitrary potential that
exists in the corresponding Newtonian solution. In the static and
locally flat limit of the Weyl class when $a>1$ it produces a linear
energy density along a string, being linked to the gravitational analog
of the Ahanarov-Bohm effect. This effect demonstrates that gravitation
depends on the topological structure of spacetime.

The parameter $b$, as we showed, is associated, in the locally flat
limit, with the angular momentum of a spinning string generating one
kind of the gravitational Aharanov-Bohm effect. 

The parameter $c$ is produced by the vorticity of the source of the Weyl 
class metric when it is represented by a general stationary completely 
anisotropic fluid. This parameter, together with the parameter $b$, is responsible
for the non-staticity of the Weyl class metric.

We proved too, using the Cartan scalars, that only $n$ helps to curve 
spacetime locally and that the three parameters $a$, $b$, $c$ only 
influence the topological structure of spacetime. As a consequence, the 
stationary Lewis metric for the Weyl class is indistinguishable locally 
from the static Levi-Civita metric as far as the curvature of spacetime is 
concerned.

\bigskip
\bigskip
\bigskip
{\bf Appendix A}

\noindent The non zero components of $R_{\mu\nu}$ for the metric
$(2.1)$ in the Van Stockum's notation [3] are $$2 e^\mu D
R^0_0=\left({lf^\prime +kk^\prime \over D} \right)^\prime,\eqno(A.1)$$
$$2 e^\mu D R^3_0= \left({fk^\prime -kf^\prime \over D}
\right)^\prime,\eqno(A.2)$$ $$2 e^\mu D R^0_3= \left( {kl^\prime
-lk^\prime \over D} \right)^\prime,\eqno(A.3)$$ $$2 e^\mu D R^3_3=
\left({fl^\prime + kk^\prime \over D}\right)^\prime,\eqno(A.4)$$
$$2R_{11}=-\mu^{\prime\prime} +\mu^\prime {D^\prime \over D} -
2{D^{\prime\prime} \over D} + {{k^\prime}^2 + f^\prime l^\prime \over
D^2},\eqno(A.5)$$ $$2R_{22}=-\mu^{\prime\prime} -\mu^\prime{D^\prime
\over D},\eqno(A.6)$$ where the primes stand for differentiation with
respect to $r$ and $$D^2=k^2 + fl.\eqno(A.7)$$ The four equations
$(A.1-4)$ are not all independent, any one of them can be expressed in
terms of remaining three.

\vfill\eject
{\bf Appendix B}

\noindent  We give below the components of the Riemann tensor in the 
standard Lorentz frame (5.4),
$$-R_{0101}=R_{2323}={1\over 4}(n^2-1)r^{-{1\over 2}(n^2+3)},\eqno(B.1)$$
$$-R_{1313}=R_{0202}={1\over 8}(n+1)(n^2-1)r^{-{1\over 
2}(n^2+3)},\eqno(B.2)$$
$$-R_{0303}=R_{1212}={1\over 8}(n-1)(n^2-1)r^{-{1\over 
2}(n^2+3)}.\eqno(B.3)$$

\bigskip
\bigskip
\bigskip
{\bf Appendix C}

\noindent The algebraic invariants of the Riemann tensor for vacuum 
spacetimes are [28]
 $R_{\alpha\beta\gamma\delta}R^{\alpha\beta\gamma\delta},$     
${R^{*\alpha\beta}}_{\gamma\delta}{R^{\gamma\delta}}_{\alpha\beta},$      
$R_{\alpha\beta\gamma\delta}R^{\gamma\delta\mu\nu}{R_{\mu\nu}}^{\alpha\beta}$,
${R^{*\alpha\beta}}_{\gamma\delta}R^{\gamma\delta\mu\nu}R_{\mu\nu\alpha\beta},
$ where $${R^{*\alpha\beta}}_{\gamma\delta}={1\over2}
{\epsilon^{\alpha\beta\mu\nu}\over \sqrt{-g}}R_{\mu\nu\gamma\delta}
\eqno (C.1)$$ 
For the Weyl class metric  
the non zero algebraic invariants using (B.1-3) are
$$R_{\alpha\beta\gamma\delta}R^{\alpha\beta\gamma\delta}={1\over 4}
(n^2 +3)(n^2 -1)^2 r^{-(n^2 +3)} ,\eqno(C.2)$$
$$R_{\alpha\beta\gamma\delta}R^{\gamma\delta\mu\nu}{R_{\mu\nu}}^{\alpha\beta}
={3\over 16}\left(n^2-1\right)^4 r^{-{3\over 2}\left(n^2+3\right)}. \eqno(C.3)$$
\bigskip
\bigskip
\bigskip
{\bf Acknowledgment}

\noindent The authors acknowledge to Bill Bonnor and Kayll Lake for reading the manuscript and for usefull comments.
\noindent MFAS and FMP gratefully acknowledge financial assistance from CAPES and CNPq, respectively.

\vfill\eject

{\bf References}
\bigskip 
\noindent
[1] Lewis T 1932 {\it Proc. Roy. Soc. Lond.} {\bf 136} 176  
\bigskip 
\noindent
[2] Kramer D, Stephani H, MacCallum M A H and Herlt E 1980 {\it Exacty 
Solutions of Einstein's Field Equations} (Cambridge: Cambridge University 
Press) p 221 
\bigskip
\noindent
[3] Levi-Civita T 1917 {\it Rend. Acc. Lincei} {\bf 26} 307  
\bigskip 
\noindent
[4] Cartan E 1951 {\it Le\c cons sur la G\'eom\'etrie des Espaces de 
Riemann} (Paris: Gauthier-Villars)
\bigskip
\noindent
[5] Karlhede A 1980 {\it Gen. Rel. Grav.} {\bf 12} 693
\bigskip
\noindent 
[6] Bonnor W B and Martins M A P 1991 {\it Class. Quantum Grav.} {\bf 8} 727 
\bigskip
\noindent
[7]Bonnor, W B 1980 {\it J. Phys. A: Math. Gen.} {\bf 13} 2121
\bigskip
\noindent
[8] Stachel, J 1982 {\it Phys. Rev. D} {\bf 26} 1281  
\bigskip
\noindent
[9] Lemos, J P S 1994 to be published
\bigskip
\noindent
[10] Krasi\' nski, A 1974 {\it Acta Phys. Polon.} {\bf B5} 411
\bigskip
\noindent
[11] Krasi\' nski, A 1975 {\it J. Math. Phys.} {\bf 16} 125
\bigskip
\noindent
[12] Krasi\' nski, A 1978 {\it Rep. Math. Phys.} {\bf 14} 225
\bigskip
\noindent
[13] Krasi\' nski, A 1975 {\it Acta Phys. Polon.} {\bf B6} 223
\bigskip
\noindent
[14] van Stockum, W J 1937 {\it Proc. R. Soc. Edin.} {\bf 57} 135
\bigskip
\noindent
[15] MacCallum M A H and Skea J E F 1994 SHEEP: a computer algebra system 
for general relativity {\it Algebraic Computing in General Relativity: 
Lecture Notes from the First Brazilian School on Computer Algebra} vol 2 ed
M J Rebou\c cas and W L Roque (Oxford: Oxford University Press)
\bigskip
\noindent
[16] MacCallum M A H 1991 Computer-aided classification of exact solutions 
in general relativity in {\it General Relativity and Gravitational Physics 
(9th Italian Conference)} ed R Cianci, R de Ritis, M Francaviglia, G Marmo,
C Rubano and P Scudellaro (Singapore: World Scientific) p 318
\bigskip
\noindent
[17] Paiva F M, Rebou\c cas M J and MacCallum M A H 1993 {\it Class. Quantum
Grav.} {\bf 10} 1165
\bigskip
\noindent
[18] Paiva F M 1993 {\it PhD Thesis} Centro Brasileiro de Pesquisas 
F\'{\i}sicas, Rio de Janeiro
\bigskip
\noindent
[19] Frick I 1977 SHEEP users guide {\it Report 77-11} Institute of 
Theoretical Physics, University of Stockholm
\bigskip
\noindent
[20] \AA man J E 1987 Manual for CLASSI-classification programs for 
geometries in general relativity (3rd provisional edn) {\it University of 
Stockholm Report}
\bigskip
\noindent
[21] MacCallum M A H and \AA man J E 1986 {\it Class. Quantum Grav.} {\bf 
3} 1133
\bigskip
\noindent
[22] Darmois E 1927 {\it M\'emorial des Sciences Math\'ematiques} (Paris: 
Gauthier-Villars) Fasc 25
\bigskip
\noindent
[23] Bonnor W B 1992 {\it Gen. Rel. Grav.} {\bf 24} 551 
\bigskip
\noindent
[24] Bonnor W B and Davidson W 1992 {\it Class. Quantum Grav.} {\bf 9} 2065  
\bigskip
\noindent
[25] Linet B 1985 {\it Gen. Rel. Grav.} {\bf 17} 1109  
\bigskip 
\noindent
[26] Dowker J S 1967 {\it Nuovo Cimento} {\bf B 52} 129 
\bigskip
\noindent
[27] Jensen B and Ku\u cera J 1993 {\it J. Math. Phys.} {\bf 34} 4975  
\bigskip
\noindent
[28] Campbell S J and Wainwright J 1977 {\it Gen. Rel. Grav.} {\bf 8} 987

\end